\documentclass[10pt,conference,letterpaper]{IEEEtran}

\usepackage{amsmath,amssymb}
\usepackage{graphicx}
\usepackage{booktabs}
\usepackage{array}
\usepackage{tabularx}
\usepackage{makecell}
\usepackage{multirow}
\usepackage{url}
\usepackage{xcolor}
\usepackage{balance}
\usepackage{cite}

\graphicspath{{Pic/}}
\newcolumntype{Y}{>{\centering\arraybackslash}X}
\newcolumntype{L}[1]{>{\raggedright\arraybackslash}p{#1}}
\usepackage[caption=false,font=footnotesize]{subfig}
\newcommand{\datasetname}{IriSig-Spoof}

\newif\ifanonymous
\anonymousfalse

\ifanonymous
\newcommand{\collectionsite}{an urban rooftop site}
\newcommand{\datasetrelease}{The repository link is omitted for double-blind review.}
\else
\newcommand{\collectionsite}{the rooftop of Xidian University in Xi'an, China}
\newcommand{\datasetrelease}{The dataset and evaluation code will be released at: \url{https://zenodo.org/records/21292280}.}

\fi

\title{\datasetname: A Real-World Benchmark for Time-Robust Satellite RF Fingerprinting and Spoofing Detection}

\ifanonymous
\author{\IEEEauthorblockN{Anonymous Authors}}
\else
\author{%
	\IEEEauthorblockN{%
		Shichang Guo\IEEEauthorrefmark{1},
		Yuanyu Zhang\IEEEauthorrefmark{1},
		Shuangrui Zhao\IEEEauthorrefmark{1},
		Ji He\IEEEauthorrefmark{1},
		Pinchang Zhang\IEEEauthorrefmark{2},
		and Yulong Shen\IEEEauthorrefmark{1}}
	\IEEEauthorblockA{%
		\IEEEauthorrefmark{1}School of Computer Science and Technology, Xidian University, Xi'an, China\\
		\IEEEauthorrefmark{2}Future University Hakodate, Hakodate, Japan\\
		Emails: scguo0117@stu.xidian.edu.cn,
		\{yyuzhang, zhaoshuangrui, jihe\}@xidian.edu.cn,\\
		zpcap0505238@163.com, ylshen@mail.xidian.edu.cn}
}
\fi
\begin{document}
\maketitle

\begin{abstract}
Low Earth orbit (LEO) satellite Internet is becoming critical communications infrastructure, yet its open wireless links remain vulnerable to satellite impersonation and signal spoofing. Radio frequency fingerprinting (RFF) offers a potential defense by exploiting transmitter-specific hardware imperfections manifested in received signals. However, the reliability of existing satellite RFF methods remains difficult to assess because no unified dataset and benchmark support temporal, open-set, and cross-scenario evaluation. To address this gap, we introduce  \datasetname, a real-world Iridium dataset comprising 5.17 million messages collected from 66 satellites over 32 days, together with software-defined radio (SDR)-generated spoofing signals from indoor and outdoor settings. We further establish three benchmark tasks: temporal robustness evaluation, open-set RFF identification with unknown-signal rejection, and cross-scenario spoofing detection. Experiments using a multi-scale attention convolutional neural network (MACNN) show that temporal robustness varies across configurations, with the best configuration achieving 97.75\% average cross-day accuracy. In open-set evaluation, MACNN achieves an area under the receiver operating characteristic curve (AUROC) of 0.9715, while showing that effective unknown-signal rejection does not necessarily ensure reliable identity assignment. Cross-scenario experiments reveal differences at low false-positive rates. \datasetname{} provides a reproducible basis for evaluating robust RFF methods under temporal variation and changing attack conditions.
\end{abstract}


\section{Introduction}
\label{sec:introduction}
Low Earth orbit (LEO) satellite constellations have become an important component of global communication infrastructure. 
They provide wide-area connectivity for maritime communications, emergency response, remote sensing, and Internet of Things services~\cite{azari2022evolution,yue2023low}. 
As these systems support increasingly critical applications, a receiver must determine whether an observed signal originated from an authorized satellite, a replay transmitter, or a ground-based spoofer~\cite{tedeschi2022satellite,jedermann2024record}. 
An accepted forged message may provide false satellite or paging information, while repeated unauthorized transmissions can interfere with the receiver's access to legitimate signaling traffic~\cite{pavur2020tale,liu2024dark}.
Cryptographic authentication remains essential for verifying message integrity and confirming that protocol traffic originates from an authorized sender. 
However, such mechanisms do not necessarily detect replayed messages, syntactically valid but unauthorized protocol traffic, or waveform-level impersonation. 

Radio frequency fingerprinting (RFF) identifies transmitters through hardware-dependent impairments introduced by their RF components~\cite{soltanieh2020review,zhang2021radio,zhaosurvey,wang2022survey}.
These impairments alter the received in-phase and quadrature (IQ) waveform and provide features that data-driven classifiers or spoofing detectors can learn~\cite{merchant2018deep,shen2023deep}. 
Recent RFF studies have used convolutional neural networks, residual networks, attention mechanisms, transformers, and self-supervised learning to improve transmitter identification~\cite{ding2018specific,oligeri2022past,deng2023lightweight,liu2023overcoming}. 

Satellite RFF differs from terrestrial settings because orbital motion, Doppler shift, changing beam assignments, and varying propagation geometries continuously alter the received waveform.
Spoofing signals may also be observed under attack environments different from those available during training.
A practical system must therefore preserve transmitter-specific discrimination over time and reject spoofing signals across propagation conditions.
Evaluation on a single date or fixed environment may substantially overestimate deployment reliability.

Existing public datasets and systems have advanced the physical-layer security of satellite Internet systems. 
PAST-AI provides a large collection of observations from the Iridium air-interface~\cite{oligeri2022past,oligeri2023physical}.
FadePrint uses the same dataset to show that fading characteristics can distinguish satellite links from terrestrial links~\cite{oligeri2024fadeprint}.
SatIQ retains high-sampling-rate IQ waveforms and evaluates resilient satellite transmitter fingerprinting~\cite{smailes2023watch}. 
A recent aerial spoofing dataset includes over-the-air attacks generated from unmanned aerial platforms~\cite{wigchert2025detection}. 
However, no unified benchmark currently combines date separation, unknown-identity separation, and multi-scenario spoofing evaluation.
As our evaluation later shows, a detector that performs nearly perfectly in a familiar environment can even reverse its decision ranking when the propagation scenario changes.

We address this gap with \datasetname, a security-oriented benchmark that integrates legitimate satellite measurements on a month-scale organized by acquisition date, multi-scenario spoofing data, and standardized evaluation procedures.
The benchmark addresses three practical security questions:
\begin{itemize}
    \item \textbf{RQ1: Temporal robustness.} How does satellite identification performance change when a fixed model is evaluated on data collected at later dates?
    
    \item \textbf{RQ2: RFF identification with unknown-signal detection.}
    Can a model trained exclusively on legitimate satellite signals identify enrolled satellites while rejecting signals from previously unobserved transmitters?
    
    \item \textbf{RQ3: Spoofing detection across scenarios.} How does the diversity of training environments affect the cross-scenario robustness of spoofing detectors?
\end{itemize}

The main contributions of this paper are as follows.

\begin{itemize}
    \item \textbf{A month-scale, date-organized satellite signal dataset.}
    We collected 5.17 million Iridium ring alert messages from all 66 operational satellites over 32 consecutive days. 
    The resulting IQ samples are synchronized and organized by acquisition date and satellite identity to support RFF identification and the evaluation of temporal robustness.

    \item \textbf{Multi-scenario physical-layer spoofing data.}
    We collected software-defined radio (SDR)-based spoofing signals in indoor and outdoor environments. 
    The resulting dataset captures different propagation conditions, enabling a systematic evaluation of spoofing detection in realistic attack scenarios.
    These data extend existing satellite RFF resources to more diverse and challenging attack conditions.

    \item \textbf{A security-oriented satellite RFF benchmark.}
    We defined standardized evaluation procedures and representative baseline configurations for temporal robustness, open-set authentication, and spoofing detection across scenarios. 
    The benchmark enables reproducible assessment under temporal variation and changes in the attack environment.
\end{itemize}

The remainder of this paper is organized as follows. 
Section~\ref{sec:background} introduces the Iridium system,  reviews satellite RFF and spoofing research, and defines the threat model. 
Section~\ref{sec:measurement} describes the signal acquisition, IQ extraction procedure, spoofing generation, quality control, and dataset organization. 
Section~\ref{sec:benchmark} defines the benchmark tasks, data partitions, and evaluation metrics. 
Section~\ref{sec:evaluation} presents the reference implementations and evaluates the three benchmark tasks. 
Section~\ref{sec:conclusion} concludes the paper.

\section{Background and Related Work}
\label{sec:background}
\subsection{Iridium User Link and Ring Alert Messages}
The Iridium constellation comprises 66 operational satellites distributed across six near-polar orbital planes.
Its user link operates in the 1616-1626.5~MHz L-band and employs a hybrid frequency division multiple access/time division multiple access (FDMA/TDMA) scheme~\cite{fossa1998overview,tan2019new}.
Each satellite contains 48 spot beams and an independent RF transmission chain whose manufacturing tolerances and nonlinearities may introduce persistent transmitter-dependent distortions.

Iridium downlink transmissions are organized into bursts~\cite{luo2023ambiguity}. 
Each burst is a short physical-layer transmission event occupying a specific carrier and TDMA time slot, while the corresponding IRA message represents the protocol-level content recovered from that burst. 
After detecting a burst, the receiver performs carrier-frequency correction, timing synchronization, demodulation, and protocol decoding. 
We focus on Iridium Ring Alert (IRA) messages because operational satellites broadcast them periodically. 
Each decoded IRA message contains a satellite identifier, which provides a reliable ground-truth label for RFF analysis.

\begin{figure}[t]
    \centering
    \includegraphics[width=0.95\columnwidth]{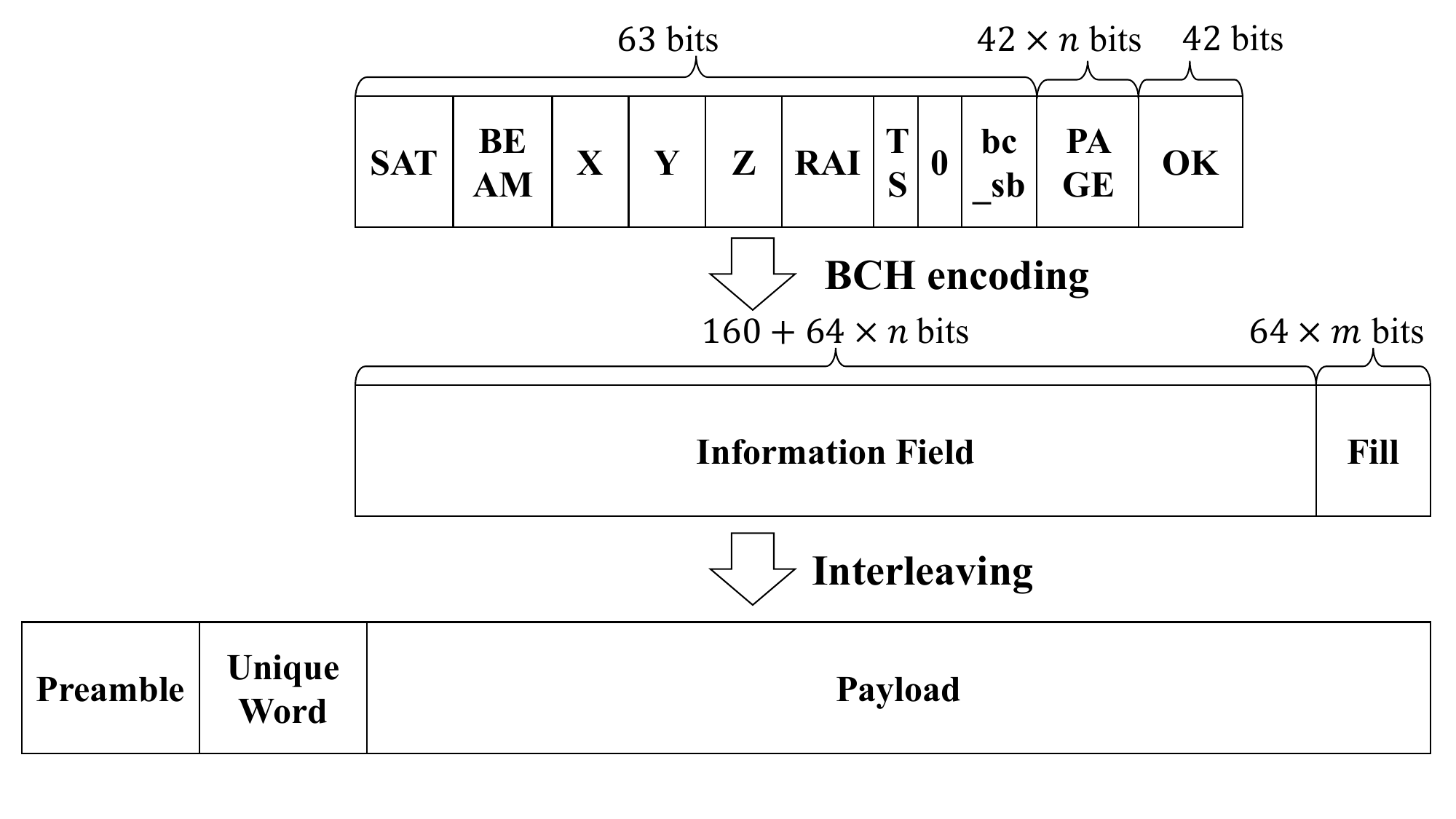}
    \caption{Logical structure of an IRA message.}
    \label{fig:IRA}
\end{figure}

As illustrated in Fig.~\ref{fig:IRA}, an IRA message contains a preamble, a unique word, and an encoded payload. 
The payload contains protocol fields that describe the transmitting satellite and its operating information. 
These fields include the satellite identifier, beam identifier, position-related fields, channel allocation, and paging information. 
Bose-Chaudhuri-Hocquenghem (BCH) coding protects the information bits, which are then combined with a predefined fill field and interleaved to form the transmitted payload. 
The interleaving operation redistributes adjacent coded bits across the message and improves the robustness to clustered transmission errors. 
The fixed preamble and unique word are combined with the encoded payload to support signal detection, synchronization, and message validation.

\subsection{Satellite RFF Datasets and Security Evaluation}
Terrestrial RFF datasets such as ORACLE and WiSig have established repeatable device-identification tasks and accelerated the development of data-driven fingerprinting~\cite{sankhe2019oracle,hanna2022wisig}. 
These datasets provide structured signal collections from multiple transmitters and support the systematic evaluation of neural network architectures, channel robustness, and device-level classification.

In contrast, publicly available satellite RFF datasets remain limited and usually focus on one aspect of satellite physical-layer analysis. 
PAST-AI provides a large-scale collection of Iridium observations and supports satellite identification from physical-layer characteristics~\cite{oligeri2023physical}. 
SatIQ preserves high-sampling-rate waveform samples and evaluates satellite transmitter fingerprinting, replay-attack detection, and temporal stability~\cite{smailes2023watch}. 
More recently, the aerial spoofing dataset introduced over-the-air attacks generated by SDR-equipped drones under different flight conditions~\cite{wigchert2025detection}.

\begin{table}[t]
\caption{Representative Satellite Physical-Layer Datasets}
\label{tab:related}
\centering
\scriptsize
\setlength{\tabcolsep}{3.2pt}
\renewcommand{\arraystretch}{1.12}
\begin{tabular}{L{0.24\columnwidth}ccc}
\toprule
\textbf{Dataset} & \textbf{Time span} & \textbf{Date-organized} & \textbf{Real spoof} \\
\midrule
PAST-AI~\cite{oligeri2023physical} & 59 days & Partial & No \\
SatIQ~\cite{smailes2023watch} & 40 days & Partial & No \\
Aerial spoofing~\cite{wigchert2025detection} & N/A & No & Yes \\
\datasetname & 32 days & Yes & Yes \\
\bottomrule
\end{tabular}
\end{table}

Table~\ref{tab:related} compares the datasets most relevant to satellite RFF authentication and spoofing detection. 
The proposed \datasetname~combines legitimate Iridium observations organized by acquisition date and satellite identity with spoofing signals collected in multiple indoor and outdoor environments. 
These spoofing measurements span different propagation conditions, so they enable a systematic evaluation of whether a detector generalizes to indoor and outdoor attack scenarios.

\subsection{Threat Model}
The adversary aims to inject false IRA messages or interfere with the receiver's access to legitimate traffic~\cite{smailes2024sticky}. 
This work focuses on detecting the unauthorized transmitting source from successfully extracted IRA waveforms rather than quantifying receiver availability under interference.
The adversary knows the publicly available IRA message structure and can use a software-defined radio to generate protocol-compliant transmissions. 
The contents of the generated messages, including the satellite identifier and other variable fields, can be configured arbitrarily. 
The adversary may place the transmitting equipment near the target receiver, allowing the injected signal to arrive with sufficient power despite using commodity radio hardware.

We do not consider precise overshadowing attacks in which the adversary synchronizes an injected waveform with a specific legitimate satellite burst and replaces it at the receiver. 
We also assume that the adversary does not compromise the legitimate satellites, the receiver hardware, or the receiver software. 
The defense objective is to identify enrolled satellites and reject IRA waveforms emitted by unauthorized nearby transmitters.

\section{Measurement and Dataset Construction}
\label{sec:measurement}
\subsection{Acquisition Platform}

\begin{figure}[t]
    \centering
    \includegraphics[width=0.95\columnwidth]{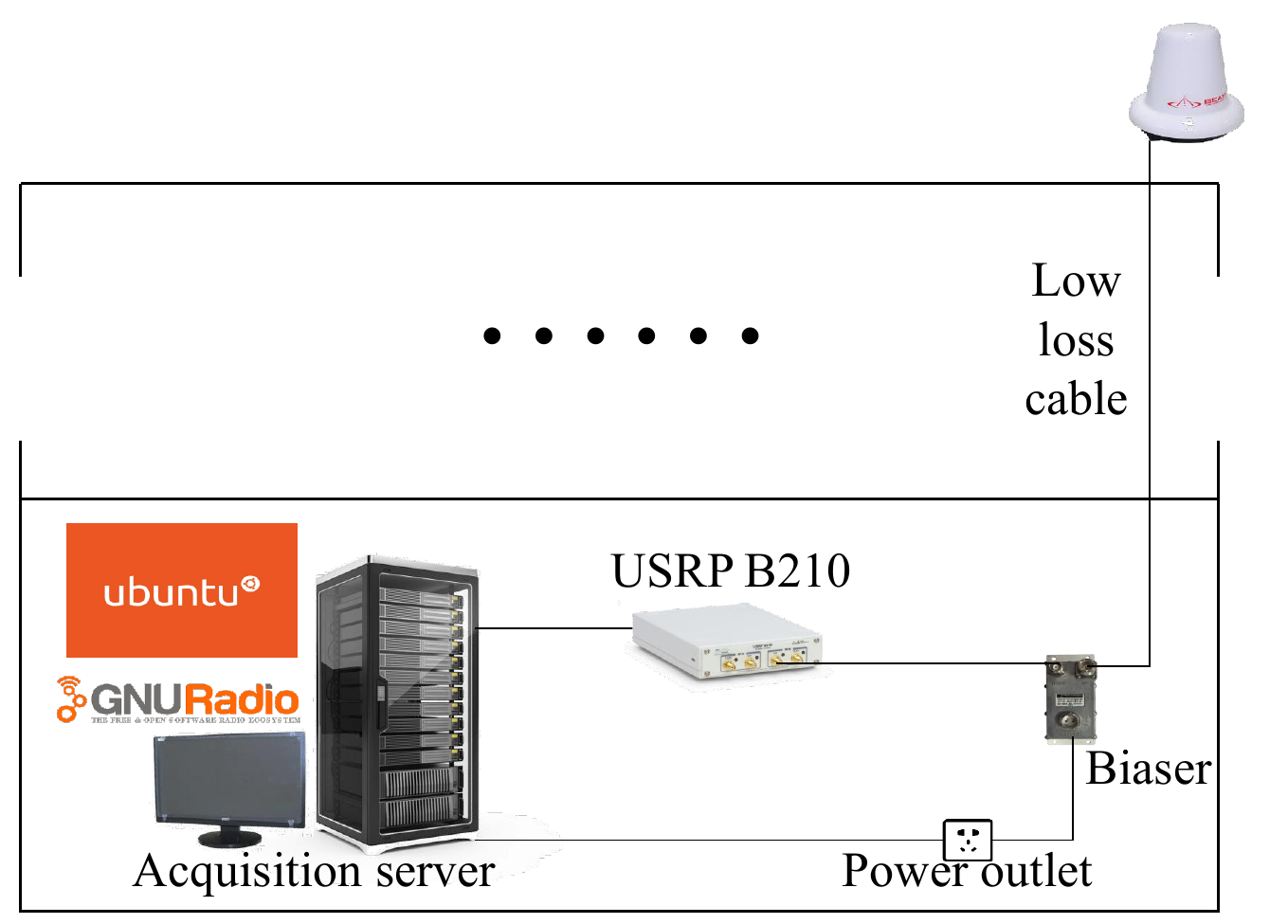}
    \caption{Signal acquisition platform used for legitimate and spoofing measurements.}
    \label{fig:platform}
\end{figure}

Fig.~\ref{fig:platform} illustrates the acquisition platform used for continuous signal collection. 
An RST740 active antenna received the L-band downlink transmissions and was powered through a bias tee over the same RF cable.
A USRP B210 performed RF downconversion and digitization to obtain complex baseband IQ samples.
The wideband acquisition front end was centered at 1.621~GHz and sampled complex baseband signals at 16~MS/s with an 8~MHz receiver bandwidth.
The receive gain was fixed at 60~dB.
The USRP transferred the samples to the acquisition server for real-time processing, decoding, and storage.
The server ran Ubuntu 22.04, UHD 4.6.0.0, GNU Radio 3.10.1.1, and a modified \textit{gr-iridium} receiver. 
The modified receiver preserved the signal segments required for the RFF analysis and associated them with decoded protocol information. 
The server used 64~GB RAM and 8~TB RAID storage for continuous acquisition.
This configuration supported long-duration acquisition and large-scale signal archiving.

\begin{figure}[t]
    \centering
    \includegraphics[width=0.95\columnwidth]{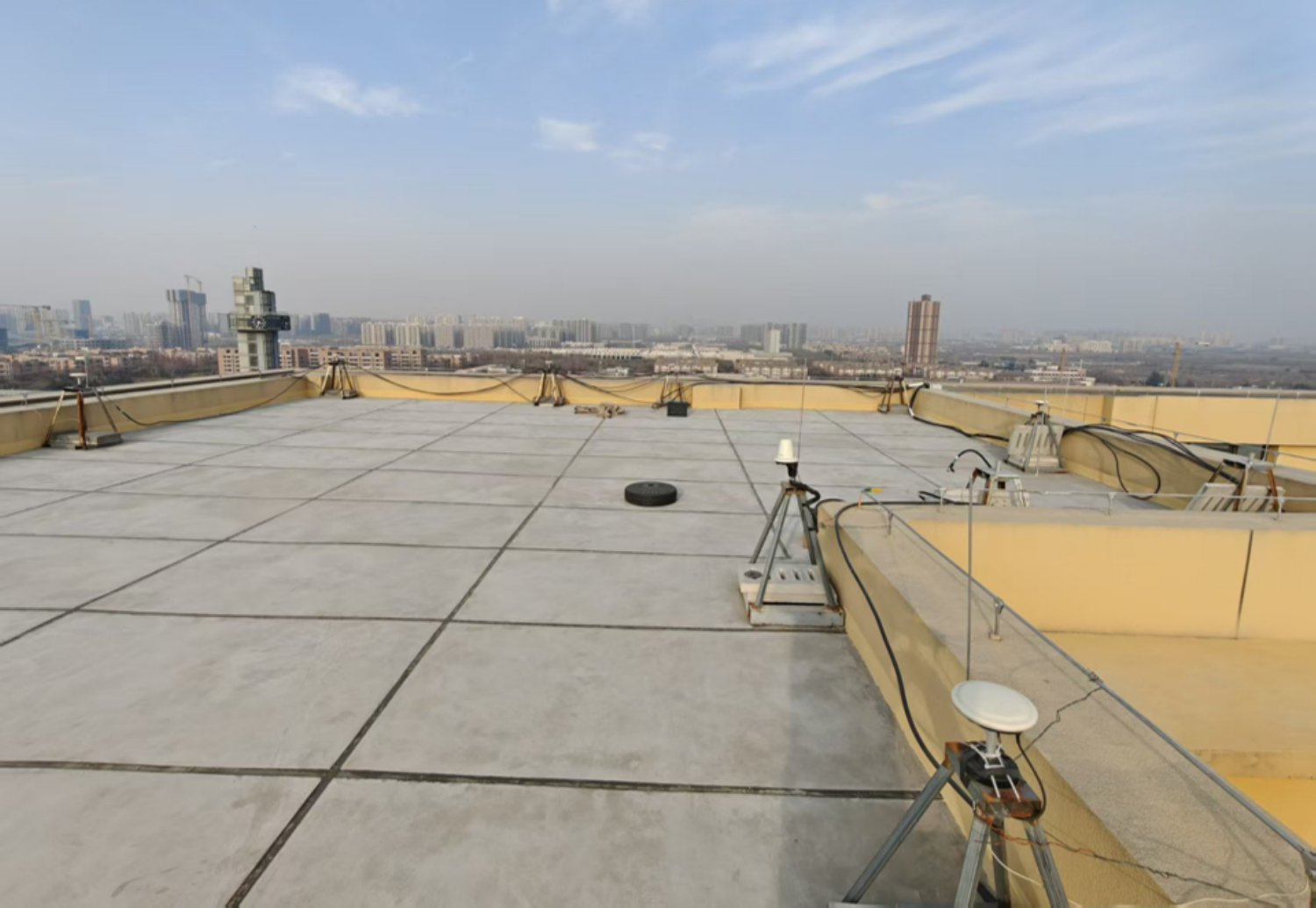}
    \caption{Deployment environment of signal acquisition antenna.}
    \label{fig:deployment}
\end{figure}

The antenna was deployed at {\collectionsite} with an unobstructed view of the sky, as shown in Fig.~\ref{fig:deployment}. 
This placement improved satellite visibility and reduced blockage from surrounding structures. 
Data collection was performed automatically in daily sessions. 
Each session started at 10:00 and ended at 09:50 on the following day, providing approximately 23~h~50~min of continuous acquisition. 
The remaining 10~min interval was reserved for file rotation, storage management, and recorder stabilization before the next session. 
The measurement campaign ran from June 27 to July 28, 2025, and covered 32 consecutive acquisition dates. 
The continuously recorded IQ logs occupied approximately 744~GB, reflecting the storage cost of long-term wideband IQ acquisition.

\subsection{Signal Extraction}
\begin{figure}[t]
    \centering
    \includegraphics[width=0.98\columnwidth]{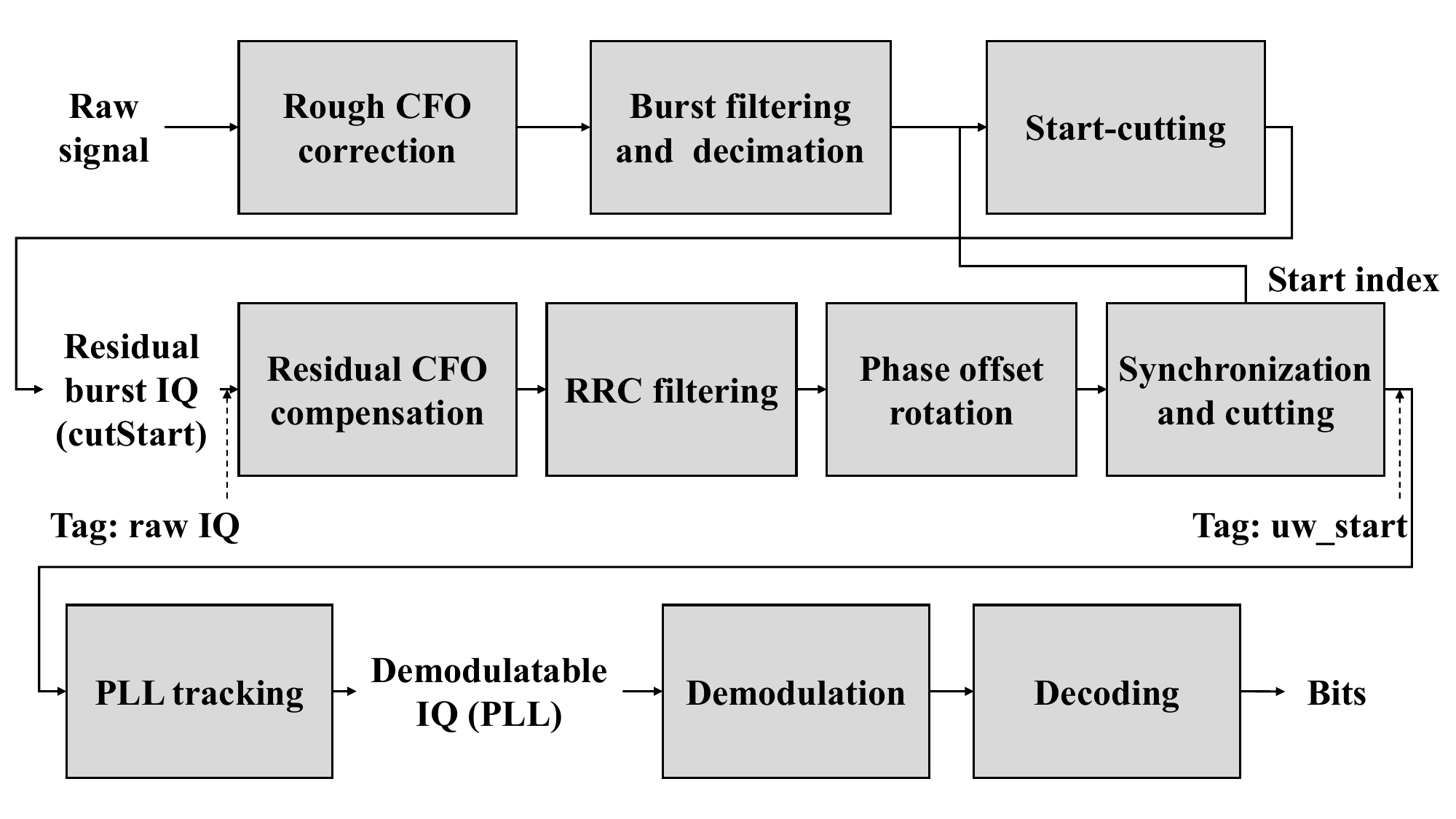}
    \caption{Modified \textit{gr-iridium} pipeline and the IQ extraction point.}
    \label{fig:pipeline}
\end{figure}

Fig.~\ref{fig:pipeline} illustrates the \textit{gr-iridium} receiver after our modifications. 
The original processing pipeline performs rough carrier frequency offset (CFO) correction, signal detection and decimation, start cutting, residual CFO compensation, root-raised-cosine (RRC) filtering, phase offset correction, synchronization, phase-locked loop (PLL) tracking, demodulation, and protocol decoding. 
These operations improve the reliability of the demodulation. 
However, the synchronization and tracking stages may suppress the frequency, phase, and symbol-transition characteristics that are useful for RFF analysis.

We introduced two data-extraction points in \textit{gr-iridium} to preserve these characteristics while retaining accurate synchronization information. 
First, we extracted the received IQ sequence immediately after the \textit{cutStart} stage. 
We stored this sequence as the raw IQ observation before residual CFO compensation, matched filtering, phase rotation, and PLL tracking. 
Then, we extended the synchronization stage to output the detected unique word start index, denoted as $uw\_start$. 
We attached the raw IQ observation and the corresponding $uw\_start$ to GNU Radio tags and exported them jointly for the construction of the dataset. 
The raw IQ observation preserved minimally processed waveform characteristics. 
After demodulation and decoding, we associated each extracted observation with the decoded satellite identity. 
This association labeled the physical-layer observation without applying the complete receiver processing pipeline to the stored IQ samples.

We jointly used the raw IQ observation and $uw\_start$ to construct each item. 
For each decoded IRA message, $uw\_start$ identifies the position of the unique word within the raw IQ observation. 
We used this position as a common synchronization reference and retained a window extending from 500 samples before $uw\_start$ to 1500 samples after it. 
We separated the resulting 2000-sample complex sequence into its in-phase and quadrature components and stored it as an array with $2\times2000$ real values. 
Each array is the standardized representation of one IRA message and constitutes a basic data unit of the complete \datasetname.

The stored samples contain received baseband waveforms rather than decoded protocol fields.
They are extracted before residual CFO compensation, matched filtering, phase rotation, and PLL tracking, and therefore retain transmitter- and reception-dependent analog variations.
Moreover, because the payload fields vary across messages and are processed through channel coding, interleaving, and differential modulation, a protocol value does not correspond to a single fixed sample-level pattern in the extracted observation.

\subsection{Spoofing Signal Generation}

\begin{figure*}[t]
    \centering
    \subfloat[Office]{
        \includegraphics[height=0.2\textwidth]{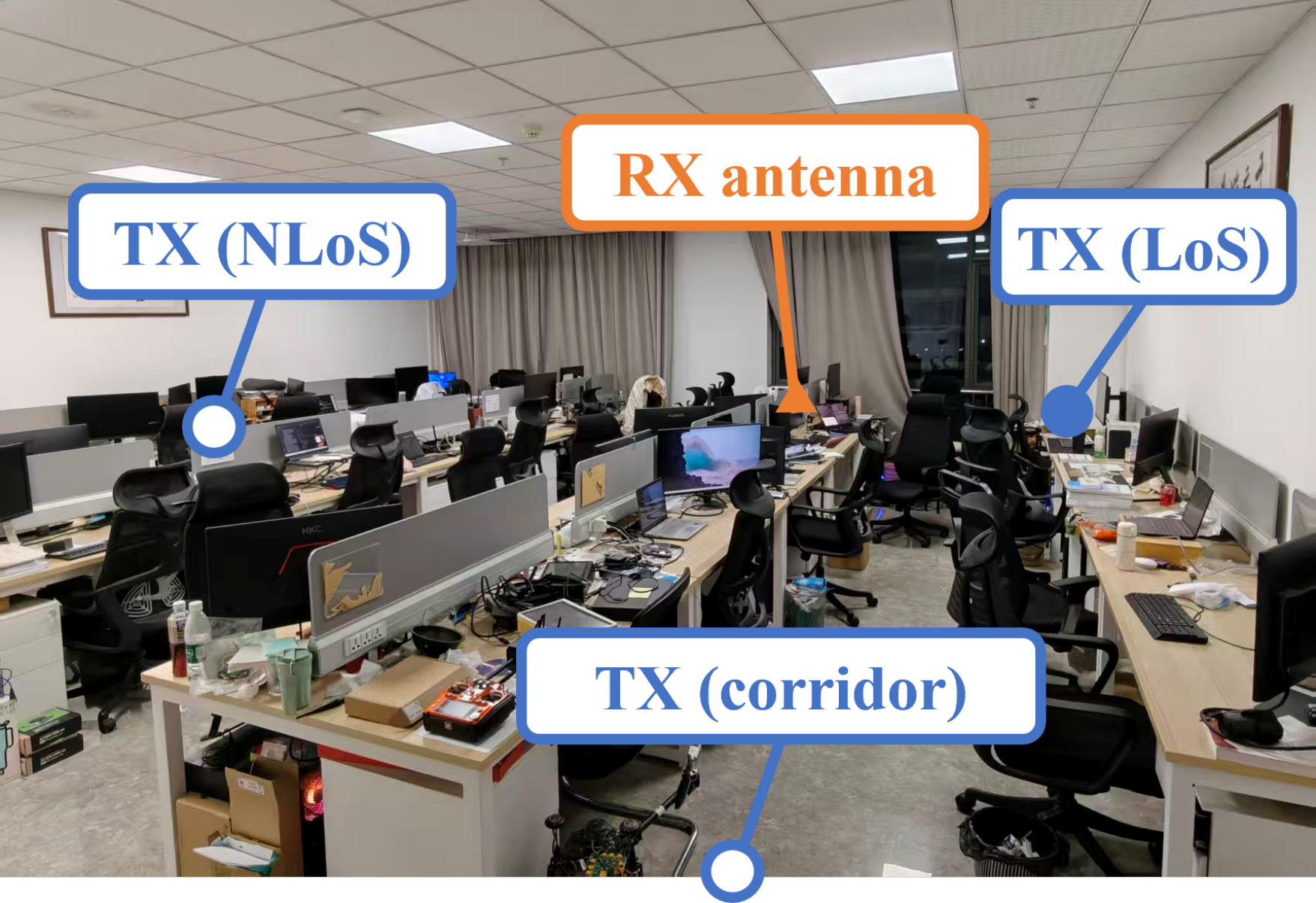}
        \label{fig:spoof_office}
    }\hfill
    \subfloat[Atrium]{
        \includegraphics[height=0.2\textwidth]{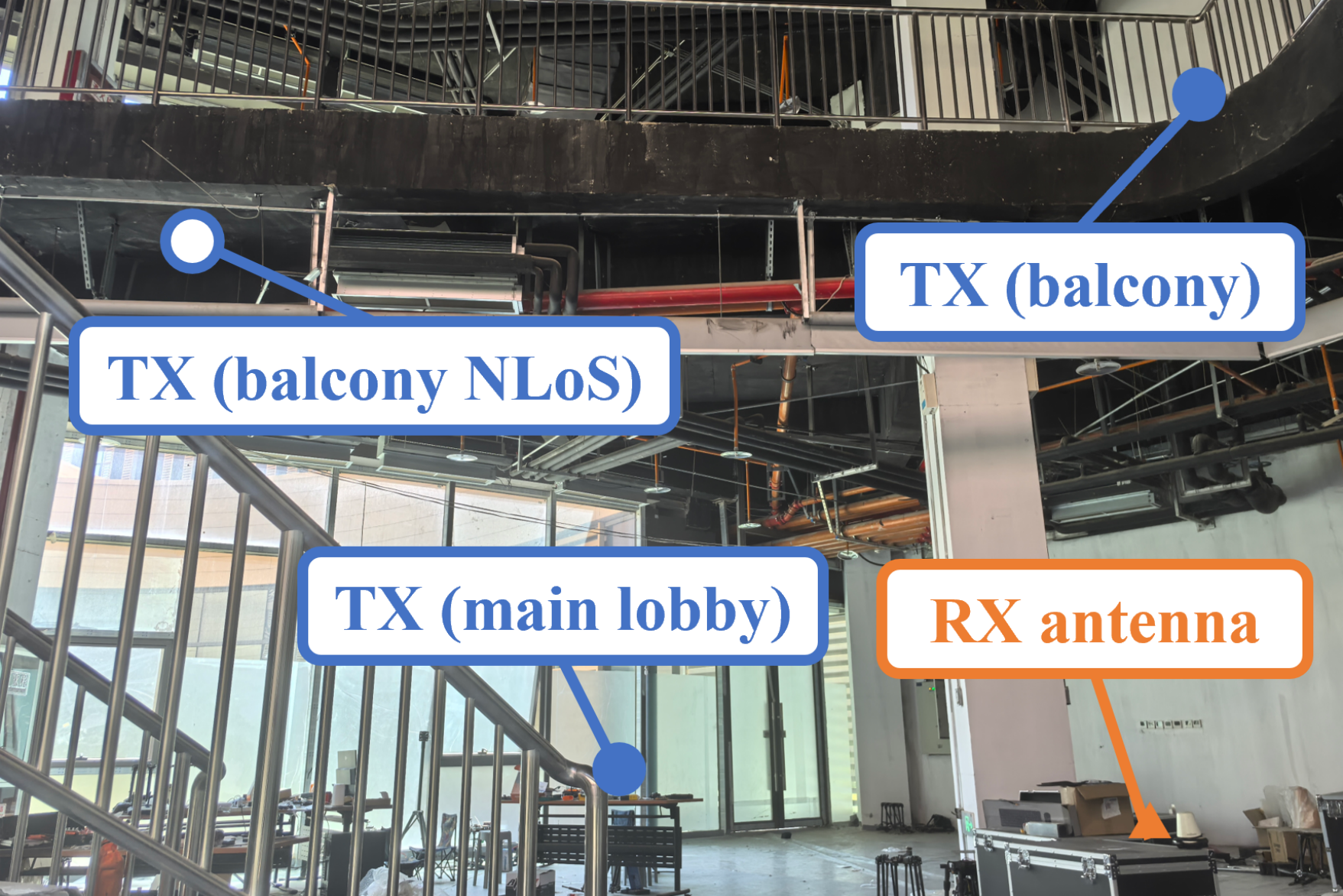}
        \label{fig:spoof_atrium}
    }\hfill
    \subfloat[Outdoor]{
        \includegraphics[height=0.2\textwidth]{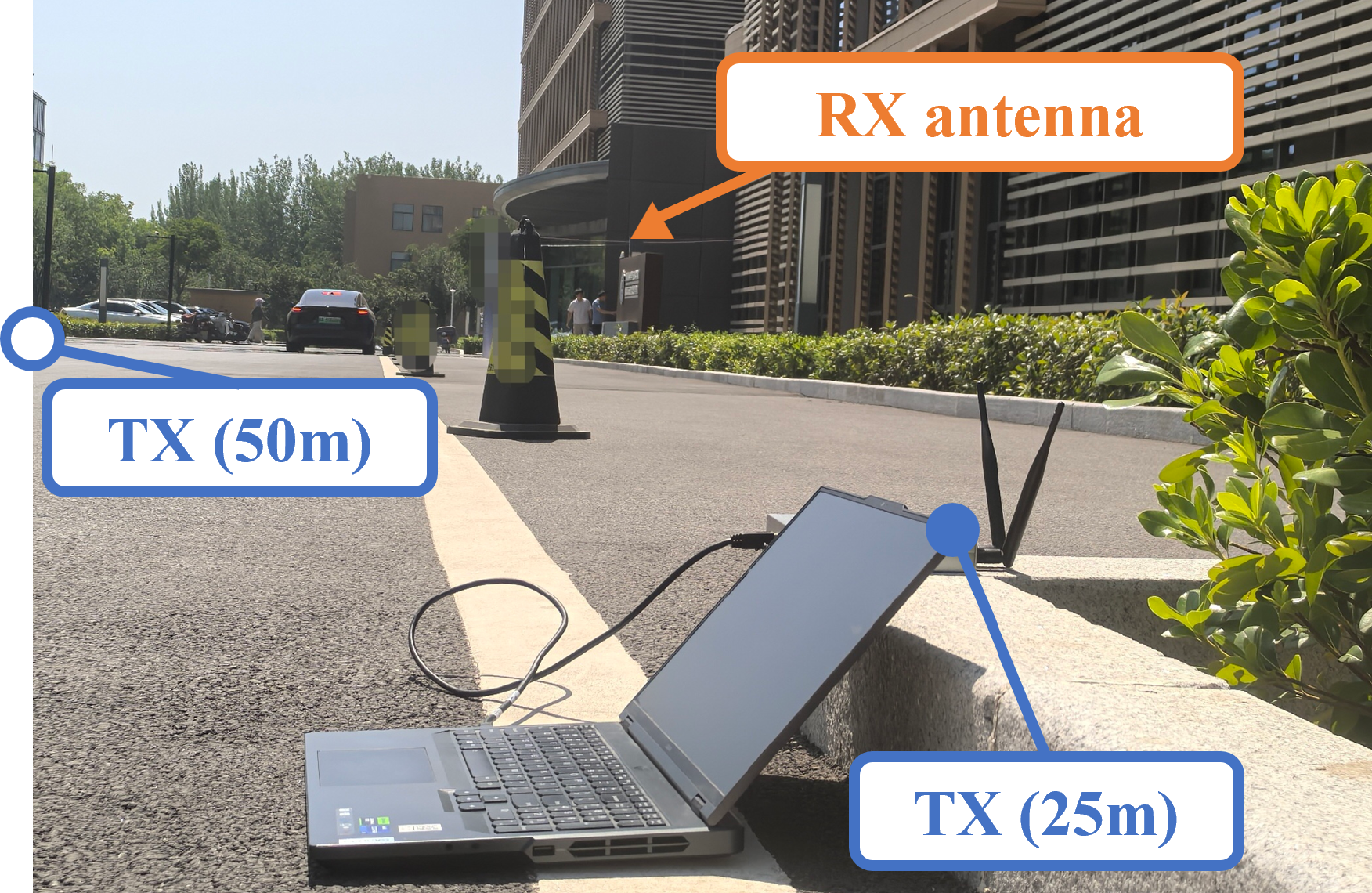}
        \label{fig:spoof_outdoor}
    }
    \caption{Spoofing-signal collection scenarios. Orange triangles indicate the receiving antenna, blue filled circles indicate visible spoofing-transmitter positions, and blue hollow circles indicate transmitters hidden behind obstacles or outside the field of view. The marked positions are schematic and were used in separate measurements rather than simultaneously.}
    \label{fig:spoofing_scenarios}
    \vspace{-2mm}
\end{figure*}

We developed an SDR-based IRA signal generator and transmitted the synthesized signals using a USRP B210. 
Each generated message followed the complete IRA construction procedure, including BCH encoding, interleaving, and differential QPSK modulation.
The position, paging, and other variable protocol fields were independently randomized for every generated message, while the satellite identifier was fixed to
$\mathrm{SAT}=1$, which is not assigned to an operational Iridium satellite.
This choice enabled unambiguous offline labeling and avoided impersonating an existing satellite during over-the-air experiments.

Although the satellite identifier is fixed, it constitutes only a small portion of the information field.
Its encoded bits are combined with the randomized message contents, redistributed by interleaving, and mapped into the differentially modulated symbols.
Consequently, the raw IQ observations do not contain a fixed local waveform segment determined solely by $\mathrm{SAT}=1$.
Instead, the symbol sequence surrounding the identifier varies across generated messages because the remaining fields and the corresponding encoded bits change from message to message.

Beyond reproducing the IRA waveform structure, the generator emulated the temporal behavior of legitimate transmissions. 
We derived the transmission schedule from the reception patterns observed during the measurement campaign and the predicted visibility of Iridium satellites over the collection site, thereby approximating the broadcast cadence of legitimate IRA traffic. 
Using this common generation pipeline, we collected spoofing signals under three representative propagation environments: an office environment, an atrium space, and an open outdoor area. 
Fig.~\ref{fig:spoofing_scenarios} illustrates the corresponding spoofing configurations.

\textbf{Outdoor scenario:} We first collected over-the-air spoofing signals in an open outdoor area at transmitter-receiver distances of 25~m and 50~m. 
This configuration more closely approximates a practical attack against an outdoor satellite terminal, where an adversary injects a competing signal from a finite distance while legitimate Iridium transmissions remain present. 
During data collection, the receiver simultaneously observed legitimate satellite traffic, environmental noise, and the spoofing signal. 
The resulting measurements therefore captured free-space path loss and uncontrolled propagation effects representative of realistic outdoor reception conditions.

\textbf{Indoor scenario:} We further collected spoofing signals in two indoor environments, an office and an atrium, to complement the outdoor measurements with a broader range of propagation conditions. 
These environments included both line-of-sight (LoS) and non-line-of-sight (NLoS) transmitter placements, together with attenuated reception of legitimate satellite signals. 
The indoor configurations introduced additional effects such as obstruction, reflection, and multipath propagation, and represent cases in which satellite terminals are temporarily operated near windows, inside public facilities, or in other semi-controlled spaces.

\subsection{Quality Control and Dataset Description}

All collected IRA records were subjected to a unified quality-control procedure before dataset construction. 
We retained a record only when the detected unique word position satisfied $uw\_start\geq500$. 
The associated raw IQ observation had to cover the complete interval from $uw\_start-500$ to $uw\_start+1500$. 
We also excluded records with metadata parsing failures. 
These criteria ensured that each retained sample could be reliably synchronized, matched with valid metadata, and represented in a uniform format.

\begin{figure}[t]
    \centering
    \includegraphics[width=\columnwidth]{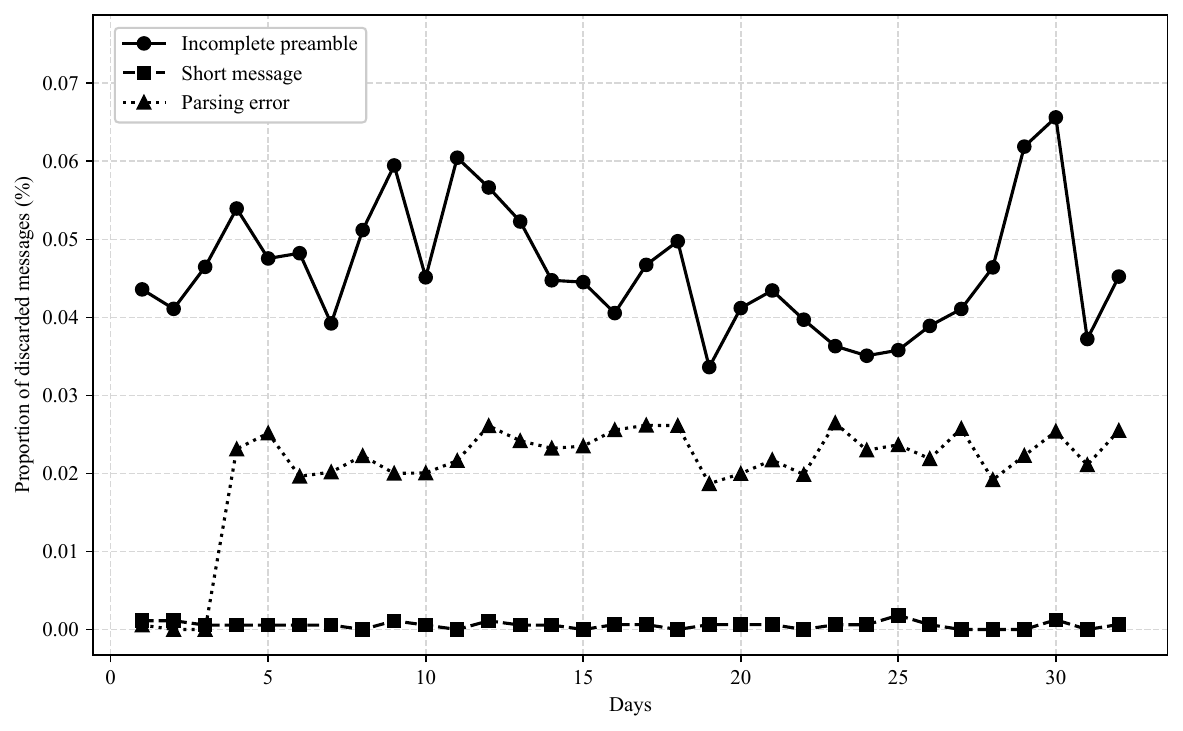}
    \caption{Numbers of IRA records discarded by the quality-control procedure across the 32 acquisition dates.}
    \label{fig:quality_control}
\end{figure}

\begin{table}
\caption{Quality-Control Results for Spoofing Signals}
\setlength{\tabcolsep}{3pt}
\begin{tabular}{p{80pt}p{65pt}p{75pt}}
\toprule
\textbf{Scenario} & \textbf{Dropped messages} & \textbf{Retained messages} \\
\midrule
Office (LoS) & 10 & 4092  \\
Office (NLoS) & 13 & 4414  \\
Office (corridor) & 16 & 4269  \\
Atrium (main lobby) & 2 & 4253 \\
Atrium (balcony) & 74 & 4166 \\
Atrium (balcony NLoS) & 75 & 4331 \\
Outdoor (25 m) & 59 & 4636  \\
Outdoor (50 m) & 26 & 3190  \\
\bottomrule
\end{tabular}
\label{tab:passrate}
\end{table}

The filtering results indicate that the collected data maintain consistently high quality throughout the measurement campaign. 
As shown in Fig.~\ref{fig:quality_control}, the proportion of the discarded legitimate records remains below 0.1\% on every acquisition date. 
Incomplete preambles are the primary cause of rejection. 
Metadata parsing errors account for a smaller and relatively stable proportion, and records with insufficient sequence lengths are rare throughout the collection period. 
The spoofing subset shows a similar quality profile. 
As summarized in Table~\ref{tab:passrate}, the quality-control procedure removes 0.8\% of the collected spoofing records. 
These results indicate that the legitimate and spoofing subsets contain sufficiently complete samples for subsequent RFF analysis.

The released \datasetname~is organized by acquisition date and satellite identity. 
Each legitimate acquisition date has a dedicated collection directory named according to the date. 
Within each directory, samples are grouped by transmitting satellite, and each file is named according to the satellite identity. 
Each file stores all valid samples from one satellite and acquisition date as an $N\times2\times2000$ real-valued array. 
In this representation, $N$ denotes the number of IRA messages, and the remaining dimensions contain the in-phase and quadrature sequences. 
The spoofing collections use the same sample representation and file format. 
Their directory names identify the attack environment or collection configuration.

\section{Benchmark Design}
\label{sec:benchmark}

The \datasetname~benchmark defines three model-agnostic evaluation tasks corresponding to the research questions introduced in Sec.~\ref{sec:introduction}: temporal robustness, RFF identification with unknown-signal detection, and spoofing detection across scenarios.
All tasks use the standardized $2\times2000$ IQ representation and control the acquisition date, transmitter identity, and spoofing configuration.

Training, validation, and test sets are separated according to the factor examined by each task.
Model selection and threshold calibration use only the training and validation sets, while the test set is reserved for final evaluation.
Table~\ref{tab:benchmark} summarizes the reference protocols and primary metrics.

\begin{table*}[t]
\caption{Proposed \datasetname~Benchmark Protocol}
\label{tab:benchmark}
\centering
\scriptsize
\setlength{\tabcolsep}{4.2pt}
\renewcommand{\arraystretch}{1.15}
\begin{tabularx}{\textwidth}
{L{0.17\textwidth} L{0.25\textwidth} X L{0.25\textwidth}}
\toprule
\textbf{Research question} &
\textbf{Training and validation} &
\textbf{Final evaluation} &
\textbf{Primary metrics} \\
\midrule

Temporal robustness &
All satellite identities from Days 1--7 &
Held-out samples from Days 1--7 and the same identities on Days 8--32, using a frozen model without adaptation &
Daily accuracy, cross-day average, standard deviation, and accuracy range \\

RFF identification with unknown-signal detection &
Legitimate signals from enrolled satellites, without spoofing samples &
Legitimate signals from enrolled satellites and spoofing signals from a transmitter absent from training &
AUROC, KAR, URR, accepted accuracy, and ACR \\

Spoofing detection across scenarios &
Balanced legitimate and spoofing signals from one or more configurations, or legitimate signals only for open-set detection &
Scenario-wise evaluation on indoor and outdoor spoofing signals &
AUROC, AUPRC, URR, and TPR@1\%FPR \\

\bottomrule
\end{tabularx}
\end{table*}

\subsection{Temporal Robustness}

This task evaluates whether an RFF model trained during an initial acquisition period remains reliable on signals collected at later dates.
All satellite identities observed during Days 1--7 are used for training and validation.
The resulting model is evaluated on held-out samples from the same period and then on each date from Days 8--32 without parameter updates, fine-tuning, or test-time adaptation.
For each date, all configurations are evaluated on the same fixed 80\% sample subset.

Daily accuracy characterizes temporal variation, while the cross-day average summarizes performance after the training period.
The standard deviation and accuracy range quantify day-to-day stability, and the difference between same-period and cross-day performance reflects sensitivity to temporal changes.

\subsection{RFF Identification with Unknown-signal Detection}

This task evaluates whether a model trained only on legitimate satellite signals can identify enrolled satellites while rejecting signals from a previously unobserved transmitter.
No spoofing samples or attacker fingerprints are used during training or validation.

For an input $x$, the model first produces a preliminary satellite identity $\hat{y}$ and an unknownness score $s_m(x)$ under score mode $m$.
The score is compared with a threshold $\tau_m$ calibrated using legitimate validation signals:
\begin{equation}
    y_{\mathrm{out}} =
    \begin{cases}
        \hat{y}, & s_m(x) \leq \tau_m,\\
        \mathrm{Unknown}, & s_m(x) > \tau_m.
    \end{cases}
\end{equation}
All scores are oriented such that a larger value indicates greater inconsistency with the enrolled satellite classes.

The area under the receiver operating characteristic curve‌ (AUROC) measures threshold-independent discrimination between legitimate and unknown signals.
The known accepted rate (KAR) and unknown rejection rate (URR) measure the acceptance of legitimate signals and the rejection of spoofing signals, respectively.
Accepted accuracy is the identification accuracy among accepted legitimate signals.
The accepted-and-correct rate (ACR) is the fraction of all legitimate signals that are both accepted and correctly identified, and is equal to the product of KAR and accepted accuracy.

\subsection{Spoofing Detection across Scenarios}

This task evaluates whether spoofing detection performance transfers across indoor and outdoor propagation conditions.
The benchmark defines supervised and open-set protocols.

The supervised protocol trains a binary detector using legitimate signals and spoofing signals from one or more collection configurations.
The legitimate and spoofing classes are balanced, and the total training-set size is fixed across configurations.
Each detector is evaluated on all spoofing scenarios, including those absent from its training data.

The open-set protocol trains an RFF model using legitimate satellite signals only.
Spoofing signals from each indoor and outdoor scenario are treated as previously unobserved transmitter signals and evaluated separately.
This protocol measures attack rejection without prior exposure to spoofing data.

Scenario-wise AUROC and area under the precision-recall curve (AUPRC) measure the overall separation between legitimate and spoofing signals.
URR reports spoofing rejection at the validation-selected threshold, while TPR@1\%FPR evaluates detection under a strict false-positive constraint.

\section{Experimental Evaluation}
\label{sec:evaluation}

\subsection{Experimental Setup}
We instantiate the three benchmark tasks using a multi-scale attention convolutional neural network (MACNN) as a common RFF backbone~\cite{chen2021multi}.
MACNN is used as a consistent reference implementation so that differences across tasks and configurations are not caused by changes in the feature-extraction backbone.

For temporal robustness, we evaluate three MACNN variants.
The first directly processes the standardized IQ samples without additional signal preprocessing or data augmentation.
The second applies the standard preprocessing pipeline before classification.
The third combines preprocessing with data augmentation and a domain-adversarial objective~\cite{wang2025avoiding} to reduce dependence on acquisition conditions.
The three variants use the same architecture and date-based evaluation protocol.

For RFF identification with unknown-signal detection, we augment the MACNN training objective with a triplet-loss term that reduces intra-satellite embedding distances and increases separation between satellite identities.
After training, legitimate samples are used to construct an embedding profile for each enrolled satellite.
The preliminary identity prediction is produced by the classifier, while the acceptance decision is obtained from classifier-confidence or embedding-based scores.
No spoofing samples are used to train the feature extractor, classifier, or satellite profiles.

For supervised spoofing detection, we replace the multiclass identification head with a binary classification head that distinguishes legitimate from spoofing signals.
Each configuration uses an equal number of legitimate and spoofing samples, and the total number of training samples is held constant.
Models trained with different combinations of spoofing configurations are then evaluated separately on all scenarios.
We additionally apply the open-set RFF model to the same scenario-wise test sets to evaluate spoofing rejection without attack data during training.

\begin{figure}[t]
    \centering
    \subfloat[Before preprocessing.]{
        \includegraphics[width=0.47\columnwidth]
        {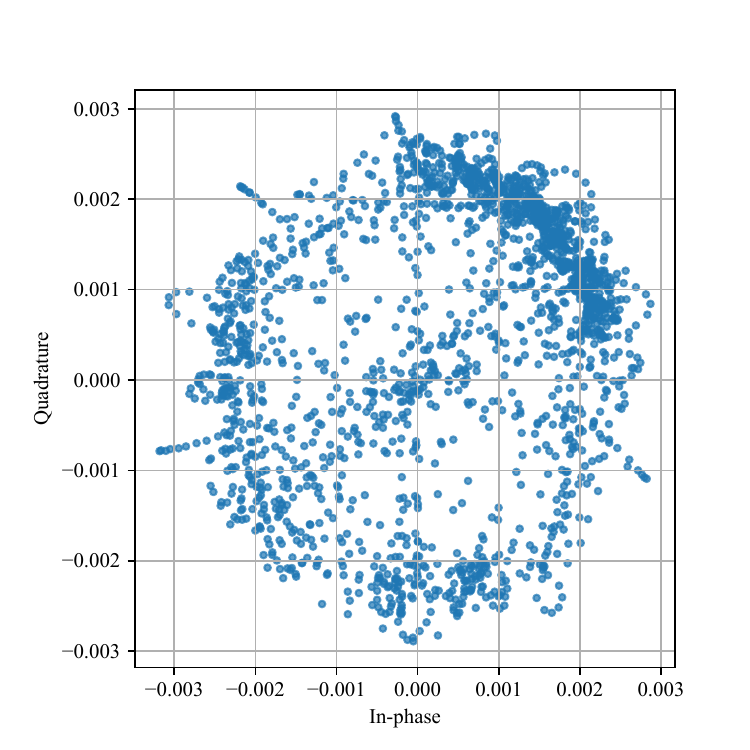}
        \label{fig:constellation-raw}
    }\hfill
    \subfloat[After preprocessing.]{
        \includegraphics[width=0.47\columnwidth]
        {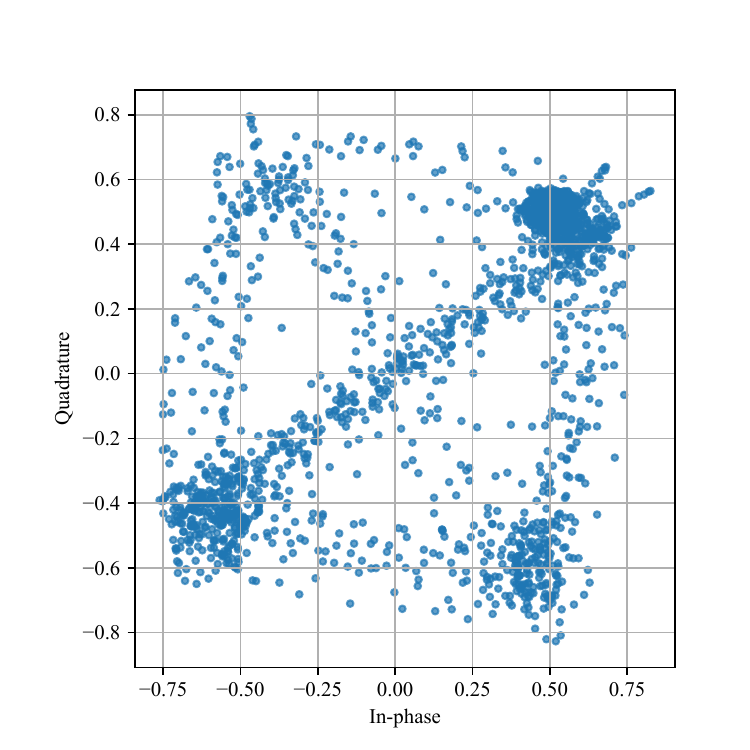}
        \label{fig:constellation-processed}
    }
    \caption{Constellation diagrams before and after standard preprocessing.}
    \label{fig:Constellation}
    \vspace{-2mm}
\end{figure}

The released dataset retains minimally processed IQ samples to preserve physical-layer characteristics observed at the receiver.
As a reproducible reference, we also provide an optional preprocessing pipeline consisting of amplitude normalization, residual CFO correction, and phase offset compensation.
Fig.~\ref{fig:Constellation} illustrates the sample distribution before and after the standard preprocessing pipeline.
All experiments use the same $2\times2000$ input representation and follow the data partitions defined in Sec.~\ref{sec:benchmark}.

\subsection{Temporal Robustness}
We evaluated three MACNN configurations on \datasetname~under the same temporal protocol. 
For every configuration, the first seven acquisition days were used for training and validation. 
After training, we froze the resulting model and evaluated it on each acquisition date without retraining. 
For each date, we randomly selected 80\% of the samples for evaluation. 

\begin{figure}[t]
    \centering
    \includegraphics[width=0.94\columnwidth]{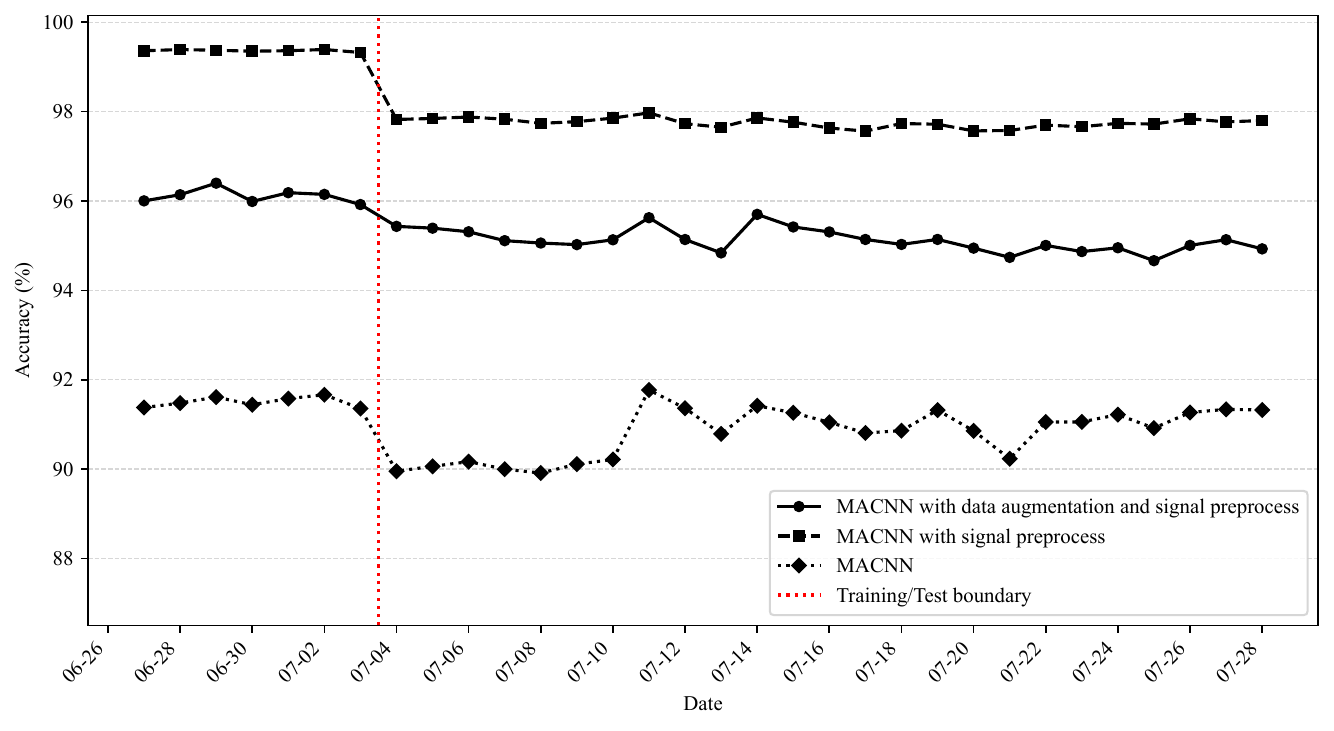}
    \caption{Cross-day accuracy of a MACNN baseline.}
    \label{fig:crossday}
\end{figure}

As shown in Fig.~\ref{fig:crossday}, all three configurations show a clear performance change after the boundary between the training and evaluation periods. 
Across the 25 dates after the training period, MACNN with signal preprocessing achieves the highest average accuracy of 97.75\%. 
Its standard deviation is 0.10 percentage points, and its total range is 0.41 percentage points.
The configuration with preprocessing and data augmentation achieves an average cross-day accuracy of 95.12\% and a standard deviation of 0.25 percentage points.
The original MACNN achieves an average accuracy of 90.81\%. 
It also exhibits the largest temporal variation, with a standard deviation of 0.55 percentage points and a range of 1.86 percentage points.

The original MACNN shows a smaller nominal gap between its in-period and cross-day average accuracies. 
However, this smaller gap primarily reflects its substantially lower in-period accuracy and does not indicate greater deployment robustness.
Under the tested protocol, the preprocessing-only configuration provides the strongest temporal robustness because it combines the highest accuracy with the lowest day-to-day variation.
The result suggests that signal preprocessing suppresses date-dependent nuisance factors while retaining satellite-specific characteristics.
By contrast, the lower accuracy of the augmented configuration indicates that the selected augmentation strategy does not represent the temporal variation of the satellite link as effectively or introduces perturbations that weaken discriminative fingerprint features.

These results show that in-period accuracy alone does not characterize long-term robustness.

\subsection{RFF Identification with Unknown-signal Detection}
The legitimate signals from the first four acquisition days are used for model training and validation. Signals collected on the fifth day from the same enrolled satellites form the known test set. The unknown test set consists of real spoofing signals collected in the outdoor 25-m and 50-m scenarios. The SDR transmitter used to generate these signals is not present in the training data. 

\begin{figure}[t]
    \centering
    \includegraphics[width=0.94\columnwidth]{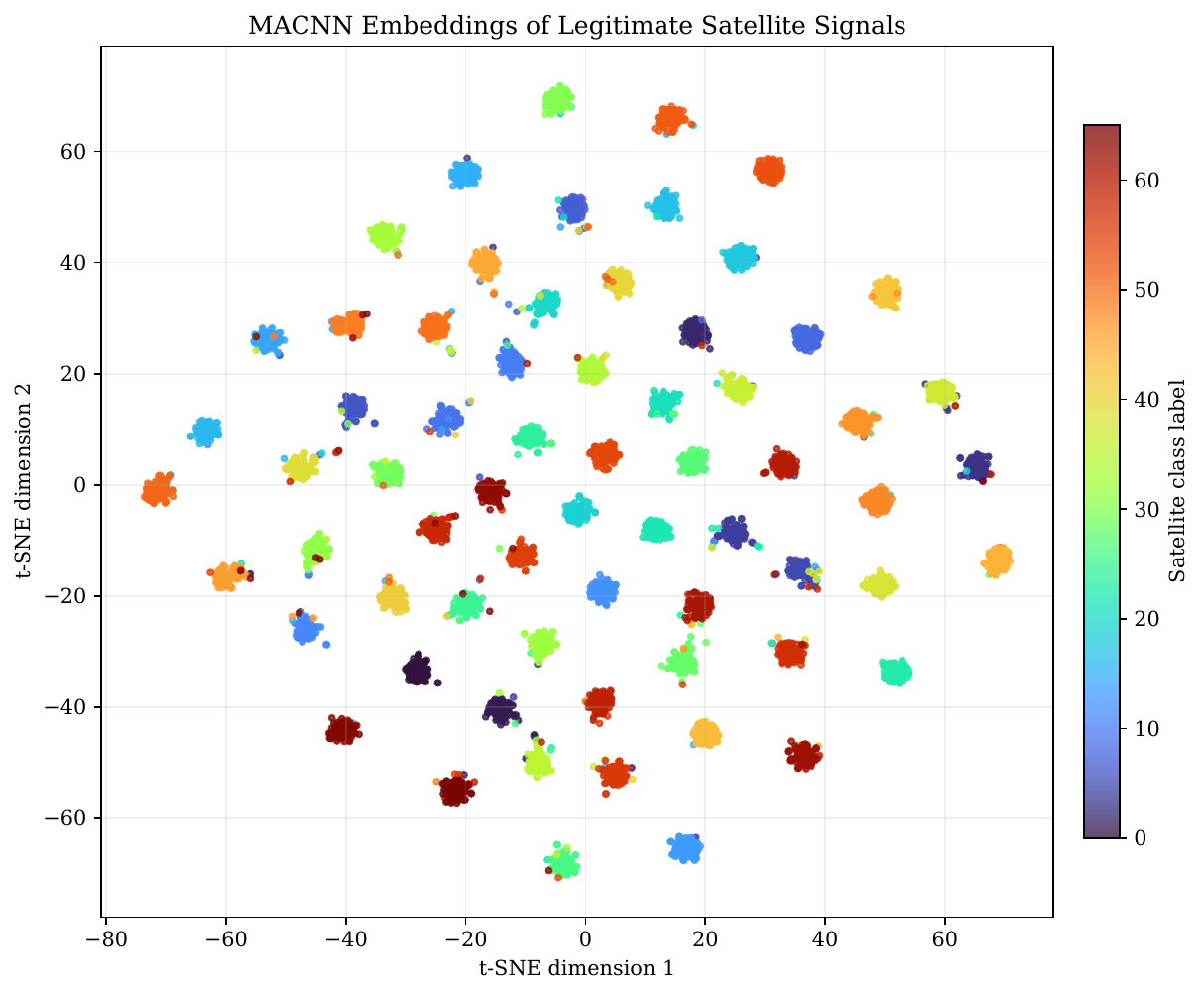}
    \caption{Embedding distribution of legitimate satellite signals.}
    \label{fig:t-sne}
\end{figure}

Fig.~\ref{fig:t-sne} presents the ‌t-distributed stochastic neighbor embedding‌ (t-SNE) projection of the embeddings of the legitimate signals.
Most satellite classes form compact and distinguishable clusters, showing that MACNN preserves satellite-dependent information in its embedding space.
A small number of overlapping points remain between neighboring clusters, which motivates evaluating identity prediction and unknown-signal rejection jointly.

We evaluate four unknownness scores using the same trained MACNN model.
All embeddings are $\ell_2$-normalized and compared using cosine distance.
Single-prototype uses the minimum distance to the mean embedding of each enrolled satellite.
Multi-proto uses the nearest of four class-specific centroids.
kNN uses the mean distance to the ten nearest embeddings in a class-balanced memory bank containing up to 128 samples per satellite.
Softmax uncertainty combines $1-p_{(1)}$ with $1-(p_{(1)}-p_{(2)})$, where $p_{(1)}$ and $p_{(2)}$ are the two largest class probabilities.
The latter three scores are standardized using the median and MAD of legitimate validation scores.
Larger scores indicate greater unknownness, and each threshold is set to the 95th percentile of its legitimate validation distribution.
No spoofing sample is used for training, profile construction, or threshold calibration.

\begin{figure}[t]
    \centering
    \includegraphics[width=0.94\columnwidth]{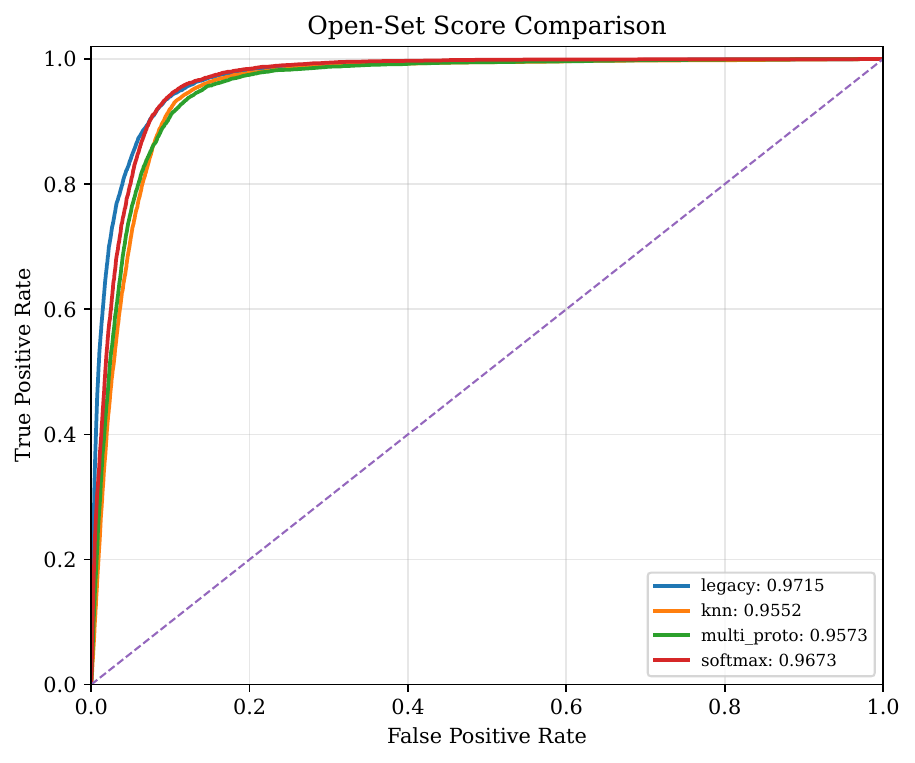}
    \caption{ROC curves of different open-set score modes.}
    \label{fig:roc}
\end{figure}

Fig.~\ref{fig:roc} compares the ROC curves produced by the four score modes. 
All configurations achieve an AUROC greater than 0.95, demonstrating that legitimate satellite signals and previously unobserved outdoor spoofing signals produce distinguishable score distributions. 
Legacy provides the strongest overall ranking, followed by Softmax, while the kNN and multi-proto scores achieve slightly lower AUROC values. 
This difference suggests that local embedding distances are more sensitive to cross-day variation than classifier confidence in this experiment. 

\begin{table}[t]
\caption{Open-Set Authentication Result}
\label{tab:openset-template}
\centering
\scriptsize
\setlength{\tabcolsep}{3.2pt}
\begin{tabular}{lccccc}
\toprule
\textbf{Score Mode} & \textbf{AUROC} & \textbf{KAR} & \textbf{URR} & \textbf{Accepted Acc.} & \textbf{ACR} \\
\midrule
Single-proto (Legacy) 
& 0.9715 & 0.9453 & 0.8583 & 0.9878 & 0.9339 \\
knn 
& 0.9552 & 0.9456 & 0.7403 & 0.9870 & 0.9334 \\
Multi-proto 
& 0.9573 & 0.9446 & 0.7777 & 0.9874 & 0.9327 \\
Softmax 
& 0.9673 & 0.9445 & 0.8325 & 0.9914 & 0.9364 \\
\bottomrule
\end{tabular}
\end{table}

Table~\ref{tab:openset-template} reveals a distinction between unknown-signal discrimination and operational satellite identification.
Legacy achieves the highest AUROC of 0.9715 and the highest URR of 0.8583, indicating the strongest separation between known and unknown signals.
Softmax, however, provides the most reliable identity decisions for legitimate signals.
It achieves the highest accepted accuracy of 0.9914 and the highest ACR of 0.9364, compared with 0.9878 and 0.9339 for Legacy, respectively.
Its KAR remains comparable to that of Legacy, decreasing only from 0.9453 to 0.9445, while its AUROC decreases by 0.0042.
These results indicate that Legacy is preferable when unknown-signal rejection is the primary objective, whereas Softmax provides a better balance when the correctness of accepted satellite identities is emphasized.

Overall, the experiment shows that \datasetname~supports joint evaluation of unknown-signal separation, legitimate-signal acceptance, and satellite identity assignment.

\subsection{Spoofing Detection across Scenarios}
We evaluate whether spoofing detectors remain reliable when the propagation environment differs from the conditions represented during training. 
The spoofing transmitter and receiver hardware remain unchanged throughout the experiment, allowing the observed differences to be attributed primarily to scenario-dependent propagation conditions.

We first consider a supervised binary detector trained using one, two, or three spoofing configurations and evaluated across all scenarios.
For a controlled comparison, every model is trained with the same number of legitimate and spoofing samples, and the total number of training samples is kept fixed across configurations.

\begin{figure*}[t]
    \centering
    \subfloat[1S, without preprocessing]{
        \includegraphics[width=0.318\textwidth]
        {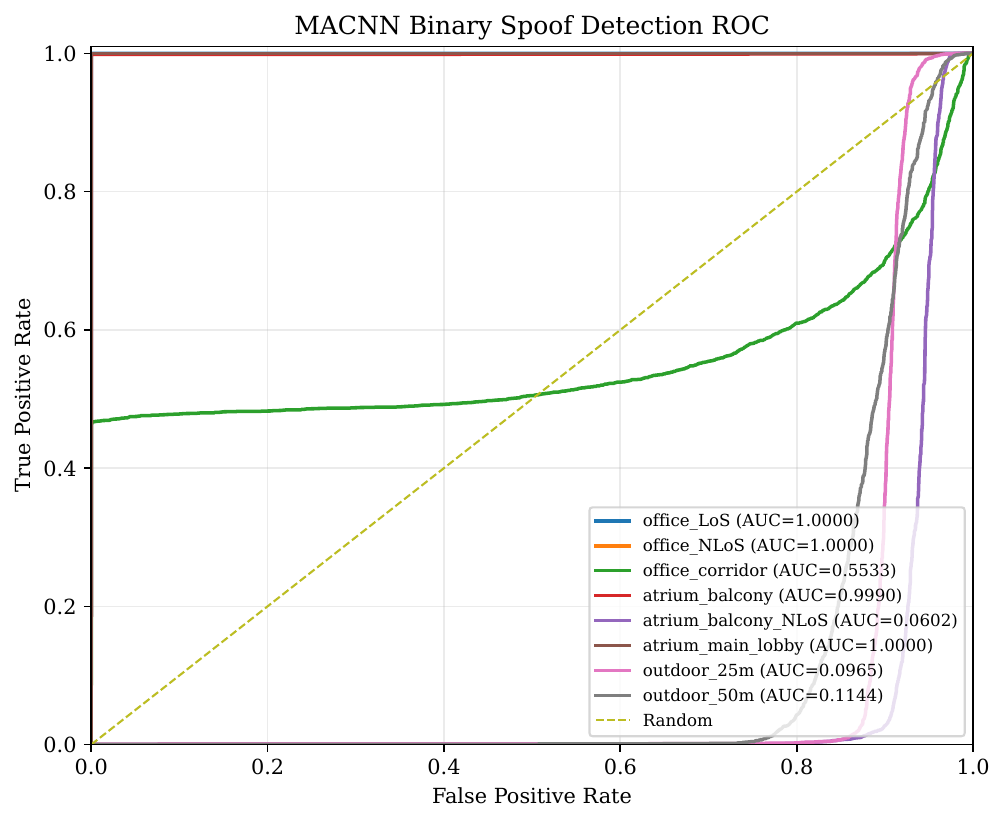}
        \label{fig:roc-1s-none}
    }\hfill
    \subfloat[2S, without preprocessing]{
        \includegraphics[width=0.318\textwidth]
        {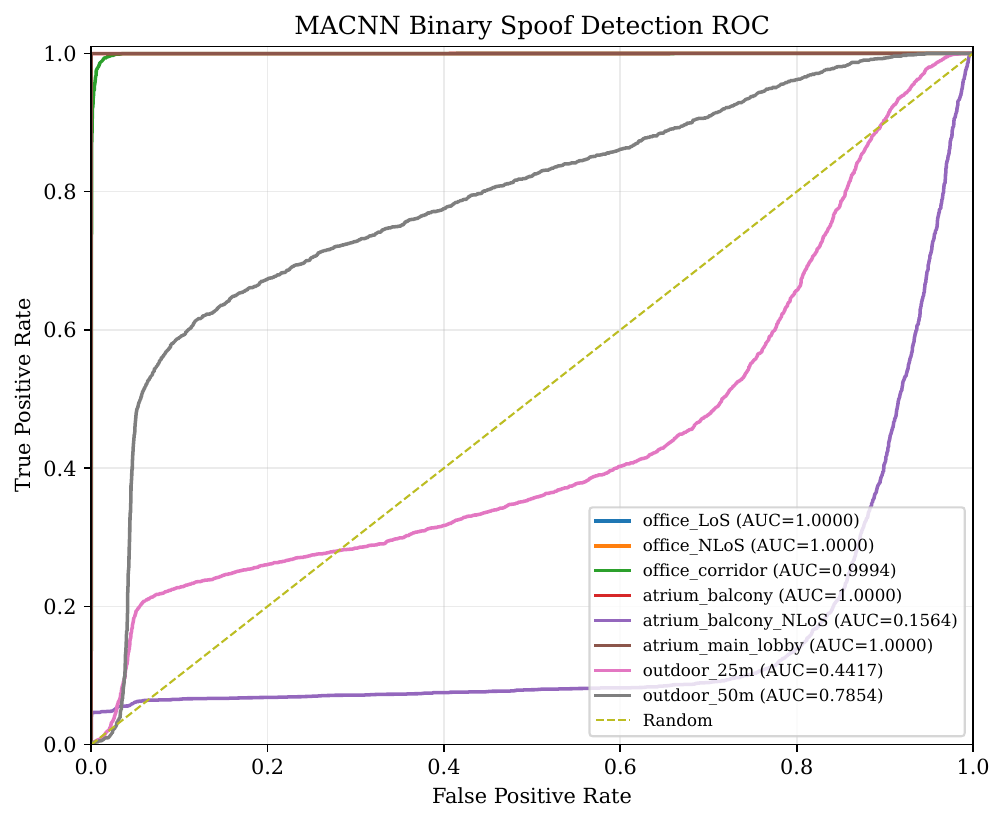}
        \label{fig:roc-2s-none}
    }
    \subfloat[3S, without preprocessing]{
        \includegraphics[width=0.318\textwidth]
        {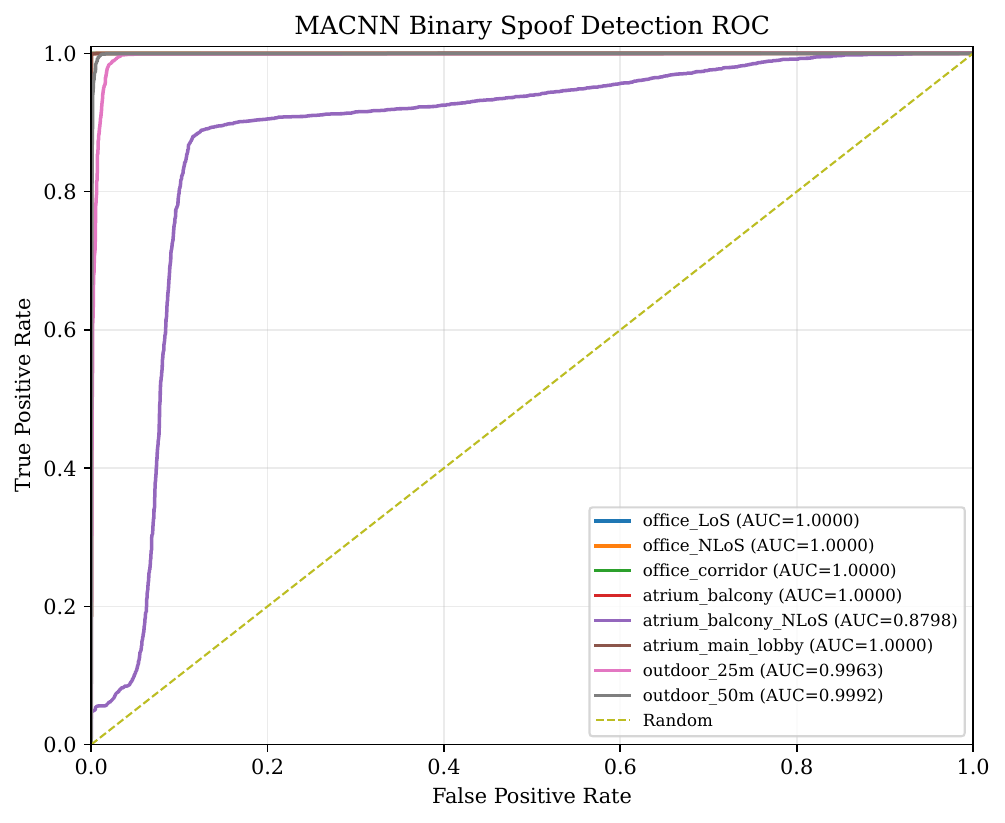}
        \label{fig:roc-3s-none}
    }\hfill

    \vspace{-1.5mm}

    \subfloat[1S, standard preprocessing]{
        \includegraphics[width=0.318\textwidth]
        {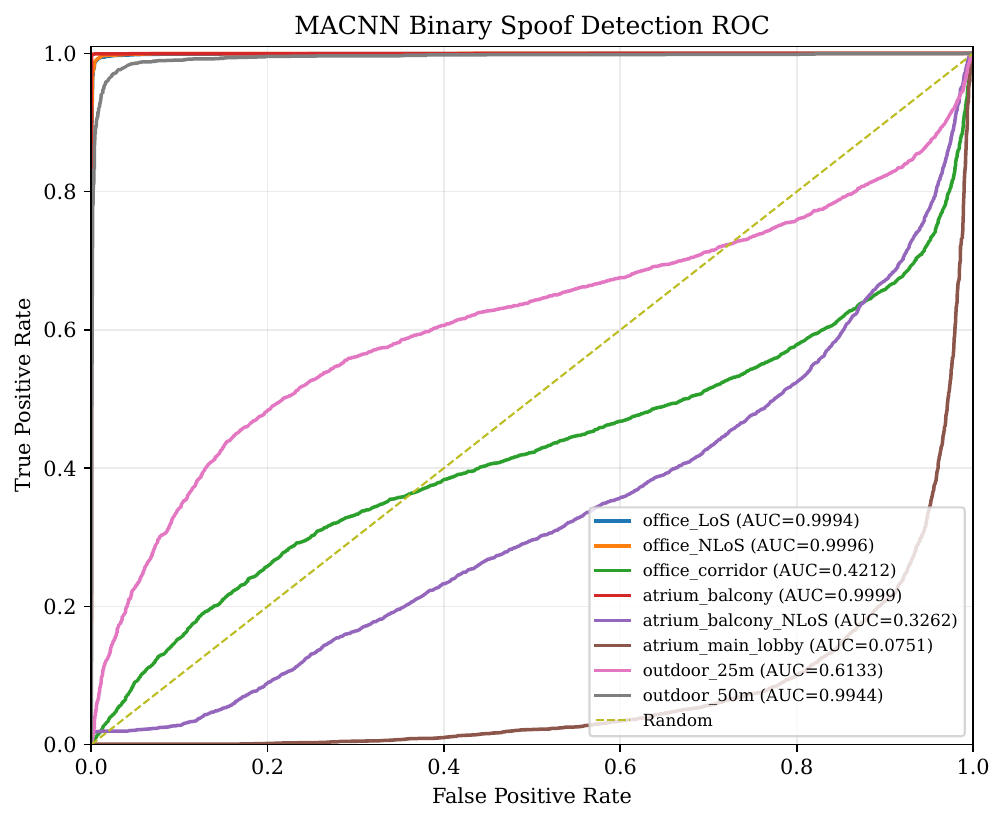}
        \label{fig:roc-1s-standard}
    }\hfill
    \subfloat[2S, standard preprocessing]{
        \includegraphics[width=0.318\textwidth]
        {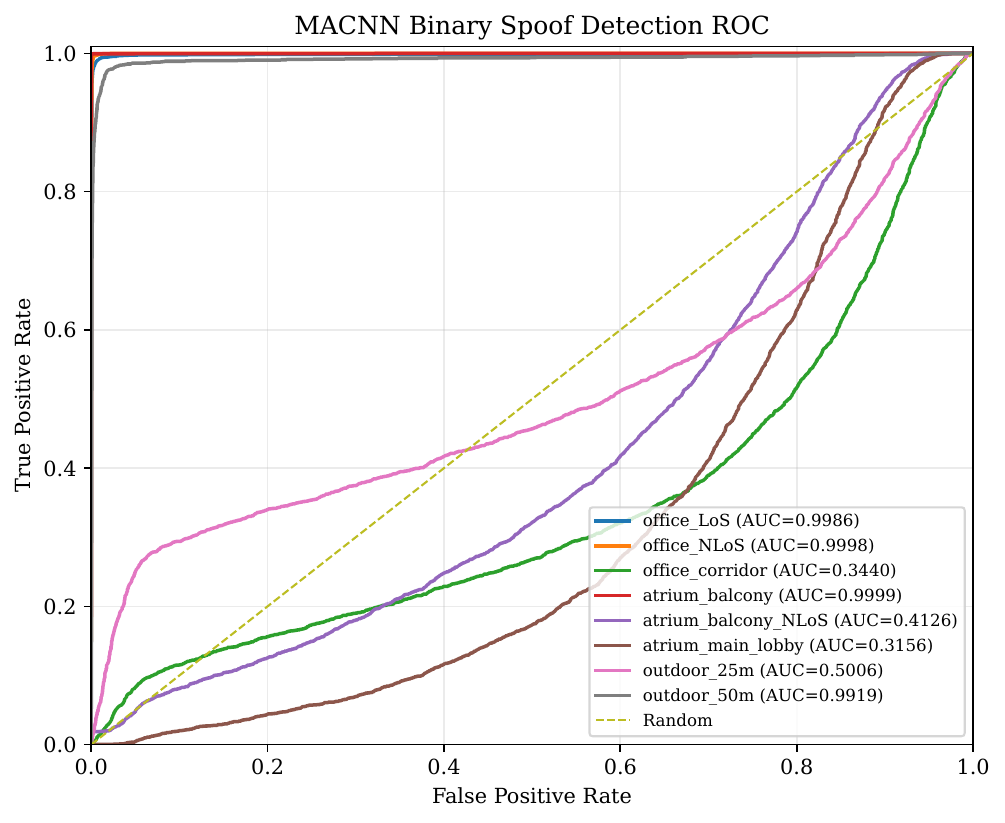}
        \label{fig:roc-2s-standard}
    }\hfill
    \subfloat[3S, standard preprocessing]{
        \includegraphics[width=0.318\textwidth]
        {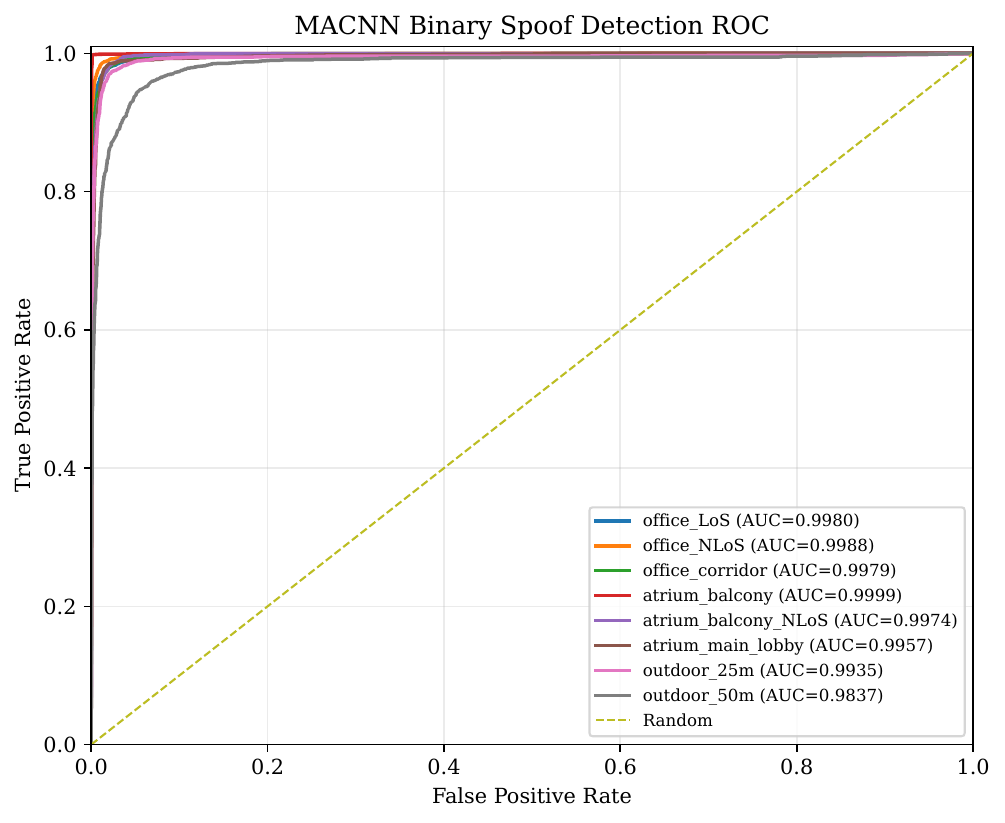}
        \label{fig:roc-3s-standard}
    }

    \vspace{-1mm}
    \caption{Cross-scenario spoofing detection under different
    training configurations. 1S uses office LoS spoofing signals,
    2S additionally includes office NLoS, and 3S further includes
    office corridor.}
    \label{fig:cross-scenario-roc}
    \vspace{-2mm}
\end{figure*}

Figure~\ref{fig:cross-scenario-roc} shows that detectors trained on one or two spoofing configurations achieve nearly perfect performance in matched scenarios but degrade substantially under unseen NLoS and outdoor conditions.
Several AUROC values fall below 0.5, indicating an inversion of the score ordering rather than random failure.
Because the transmitter, receiver, and signal generator remain fixed, this behavior is consistent with the model exploiting propagation-dependent shortcuts such as amplitude, obstruction, and multipath characteristics instead of environment-invariant transmitter features.

Increasing the training diversity from 1S to 3S produces consistently high AUROC across the remaining indoor and outdoor scenarios.
Standard preprocessing does not remove the scenario dependence of the 1S and 2S models, whereas broader training coverage improves both preprocessing settings.
These results indicate that scenario diversity contributes more reliably to cross-scenario robustness than preprocessing alone.

\begin{table}[t]
\caption{Cross-Scenario Spoofing Result}
\label{tab:scenario-template}
\centering
\scriptsize
\setlength{\tabcolsep}{2.4pt}
\begin{tabular}{lcccc}
\toprule
\textbf{Scenario} & \textbf{AUROC} & \textbf{AUPRC} & \textbf{URR} & \textbf{TPR@1\%FPR} \\
\midrule
Office (LoS)      
& 0.9695 & 0.8991 & 0.8626 & 0.5924 \\
Office (NLoS)     
& 0.9705 & 0.9052 & 0.8650 & 0.5837 \\
Office (corridor) 
& 0.9683 & 0.8956 & 0.8566 & 0.5589 \\
Atrium (main lobby)
& 0.9693 & 0.9013 & 0.8650 & 0.5874 \\
Atrium (balcony)
& 0.9672 & 0.8956 & 0.8591 & 0.5787 \\
Atrium (balcony NLoS)
& 0.9709 & 0.9051 & 0.8739 & 0.5821 \\
Outdoor (25m) 
& 0.9694 & 0.8847 & 0.8605 & 0.5392 \\
Outdoor (50m) 
& 0.9714 & 0.8312 & 0.8546 & 0.5194 \\
\bottomrule
\end{tabular}
\end{table}

The open-set RFF model is trained only on legitimate satellite signals, while signals from the spoofing transmitter are treated as previously unobserved sources during evaluation. 
Table~\ref{tab:scenario-template} reports its performance separately for each spoofing scenario.
The AUROC remains close to 0.97 and the URR remains around 0.86 across all scenarios, indicating stable overall separation between legitimate satellites and the previously unobserved spoofing transmitter.
However, AUPRC and TPR@1\%FPR decrease in the outdoor scenarios, particularly at 50~m.
This result shows that strong threshold-independent ranking does not necessarily provide equally reliable detection at a strict operating point.

The supervised detector reveals how the diversity of available spoofing data affects performance across scenarios, whereas the open-set detector evaluates rejection without prior exposure to the attacker.
The scenario-wise results demonstrate that \datasetname~can expose performance differences that would be hidden by same-scenario or aggregated evaluation.

\section{Conclusion}
\label{sec:conclusion}

We presented \datasetname, a month-scale Iridium dataset and benchmark organized by date, spoofing scenario, and satellite identity.
The results show that preprocessing affects temporal robustness, unknown-signal ranking does not directly translate to reliable identity assignment, and detectors trained on limited scenarios may learn propagation-dependent shortcuts.
Increasing scenario diversity improves cross-scenario transfer, highlighting the need to evaluate satellite physical-layer security under both time and attack conditions.

The dataset uses one spoofing SDR device and one primary receiver pipeline over approximately one month.
Therefore, it does not capture seasonal drift or generalization across attacker hardware, receiver configurations, collection sites, and more advanced attack conditions.
Future work will extend the duration of collection, hardware diversity, and attack environments.

\datasetrelease

\balance
\bibliographystyle{IEEEtran}
\bibliography{references}

\end{document}